\documentclass[runningheads]{llncs}
\usepackage{amsmath}
\usepackage{amssymb}
\usepackage[matrix,arrow,curve]{xy}
\usepackage{etoolbox}
\usepackage{thmtools}
\usepackage{tikz}
\usetikzlibrary{positioning}

\usepackage{color}

\newbool{CONFVER}
\booltrue{CONFVER}
\boolfalse{CONFVER}

\ifbool{CONFVER}{}{
\usepackage{hyperref}
}

\usepackage{cleveref}

\newcommand{\supp}{\mathrm{supp}}
\newcommand{\tin}{\mathrm{in}}
\newcommand{\tout}{\mathrm{out}}
\newcommand{\CCout}{\mathrm{out}}
\newcommand{\tgr}{\mathcal{R}}

\newenvironment{Proof}{\begin{proof}}{\qed\end{proof}}

\ifbool{CONFVER}{
\title{Dominance and T-invariants for Petri Nets and Chemical Reaction Networks}
}{
\title{Dominance and Deficiency for Petri Nets and Chemical Reaction Networks}
}
\author{Robert Brijder}
\institute{Hasselt University and Transnational University of Limburg, Belgium \email{robert.brijder@uhasselt.be}}

\begin{document}

\maketitle

\begin{abstract}
Inspired by Anderson et al. [J.\ R.\ Soc.\ Interface, 2014] we study the long-term behavior of discrete chemical reaction networks (CRNs). In particular, using techniques from both Petri net theory and CRN theory, we provide a powerful sufficient condition for a structurally-bounded CRN to have the property that none of the non-terminal reactions can fire for all its recurrent configurations. We compare this result and its proof with a related result of Anderson et al. and show its consequences for the case of CRNs with deficiency one.
\end{abstract}

\section{Introduction}
Chemical reaction network (CRN) theory studies the behavior of chemical systems. Traditionally, the primary focus is on continuous CRNs, where mass action kinetics is assumed, see, e.g., \cite{ArisCRN,Feinberg/deficiency,FeinbergHorn/deficiency,Horn/deficiency}. In this setting a state is determined by the concentration of each species and the system evolves through ordinary differential equations. However, in scenarios where the number of molecules is small one needs to resort to discrete CRNs. In a discrete CRN a state (also called configuration) is determined by the counts of each species, and one often associates a probability to each reaction. In this paper we consider only discrete CRNs, and so, from now on, by CRN we will always mean a discrete CRN.

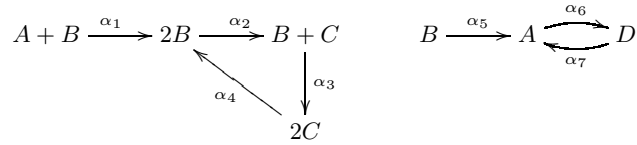
\begin{figure}
\begin{center}
$
\xymatrix{
A+B \ar[r]^{\alpha_1} & 2B \ar[r]^{\alpha_2} & B+C \ar[d]^{\alpha_3} & B \ar[r]^{\alpha_5} & A \ar@/^5pt/[r]^{\alpha_6} & D \ar@/^5pt/[l]^{\alpha_7}\\
& & 2C \ar[ul]^{\alpha_4}
}
$
\end{center}
\caption{The reaction graph of a CRN $N$.}
\label{fig:ex_intro1}
\end{figure}

A CRN essentially consists of a finite set of reactions such as $A+B \to 2B$, which means that during this reaction one molecule of species $A$ and one molecule of species $B$ are consumed and as a result two molecules of species $B$ are produced. We may depict a CRN as a graph, the reaction graph, where the vertices are the left-hand and right-hand sides of reactions and the edges are the reactions, see Figure~\ref{fig:ex_intro1} for an example. We focus in this paper on the long-term behavior of CRNs for which the number of molecules cannot grow unboundedly. For such CRNs, called structurally-bounded CRNs, each configuration eventually reaches a configuration $c$ such that $c$ is reachable from any configuration $c'$ reachable from $c$ (i.e., we can always go back to $c$). Such configurations are called recurrent. The CRN $N$ of Figure~\ref{fig:ex_intro1} is structurally-bounded.

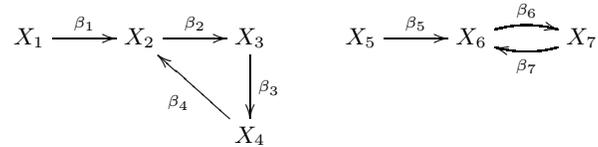
\begin{figure}
\begin{center}
$
\xymatrix{
X_1 \ar[r]^{\beta_1} & X_2 \ar[r]^{\beta_2} & X_3 \ar[d]^{\beta_3} & X_5 \ar[r]^{\beta_5} & X_6 \ar@/^5pt/[r]^{\beta_6} & X_7 \ar@/^5pt/[l]^{\beta_7}\\
& & X_4 \ar[ul]^{\beta_4}
}
$
\end{center}
\caption{The reaction graph of the CRN $N'$ obtained from $N$ by introducing a distinct species $X_i$ for each vertex.}
\label{fig:ex_intro2}
\end{figure}

Now, let us consider the CRN $N'$ obtained from $N$ by replacing every vertex by one molecule of a distinct species $X_i$, see Figure~\ref{fig:ex_intro2}. We easily observe that for $N'$, the recurrent configurations are exactly those without molecules of species $X_1$ or $X_5$. In other words, the reactions $\beta_1$ and $\beta_5$ cannot fire for any recurrent configuration of $N'$. Notice that the reaction graph of $N'$ has two strongly-connected components without outgoing edges: one having the vertices $X_2$, $X_3$, and $X_4$ and one having the vertices $X_6$ and $X_7$. The reactions outside these two strongly-connected components are called non-terminal. Thus $N'$ has the property that none of the non-terminal reactions can fire for all its recurrent configurations. But what about the original CRN $N$? The dynamics of $N$ are clearly more involved since we can go, for example, from configuration $A+B$ back to $A+B$ by firing reaction $\alpha_1$ followed by firing reaction $\alpha_5$.

%If $N$ would have deficiency zero (a well-studied measure use in CRN theory, see Subsection~\ref{ssec:defic}), then, similar as for $N'$, none of the non-terminal reactions can fire for all its recurrent configurations.

The main result of this paper, cf.\ \Cref{thm:main_result}, is a sufficient condition for a structurally-bounded CRN to have the property that none of the non-terminal reactions can fire for all its recurrent configurations (we recall the notion of non-terminal reaction in \Cref{sec:main_result}). Those CRNs have relatively simple long-term behavior. The sufficient condition of \Cref{thm:main_result} (when formulated in terms of so-called T-invariants in \Cref{cor:main_result}) is structural/syntactical and can be checked for many CRNs in a computationally-efficient way. Various non-trivial CRNs from the literature satisfy the sufficient condition of \Cref{thm:main_result} (see, e.g., the CRNs given in \cite{Anderson/CRN/Domination}), and so it can make non-trivial predictions about the long-term behavior of those CRNs. In particular, the CRN $N$ of Figure~\ref{fig:ex_intro1} satisfies the sufficient condition. Moreover, this result can also be used as a tool for engineering CRNs that perform deterministic computations (independent of the probabilities), such as in the computational model of \cite{DBLP:conf/dna/ChenDS12}. Indeed, such CRNs generally require relatively simple long-term behavior which may be partially verified by \Cref{thm:main_result}.

%\corr{MEER INTUITIE: deficiency zero is replace with unique vertices, terminal are the only reaction than that can fire in the long term states (=recurrent states for structurally bounded), but this can also happen for higher deficiency networks.}

\Cref{thm:main_result} is inspired by the main technical result of \cite{Anderson/CRN/Domination} (which in turn was inspired by the main result of \cite{Shinar12032010}), which provides another sufficient condition for the non-applicability of non-terminal reactions for recurrent configurations. However, there are a number of differences between both results. First, \Cref{thm:main_result} is derived in a basic combinatorial setting using notions from Petri net theory such as the notion of T-invariant, without considering stochastics. In contrast, the intricate proof of the main result of \cite{Anderson/CRN/Domination} is derived in a very different setting that uses non-trivial arguments from both mass action kinetics and stochastics. Secondly, we show examples where the main result of \cite{Anderson/CRN/Domination} is silent, while \Cref{thm:main_result} makes a prediction. In fact, we conjecture that the main result of \cite{Anderson/CRN/Domination} is a special case of \Cref{thm:main_result}. We compare both results in detail in \Cref{sec:using_rates}. While we focus in this paper on recurrent configurations of CRNs, we mention that the related concept of recurrent CRN has been investigated in \cite{DiscCRN:LCNrecurr}.

\ifbool{CONFVER}{}{Deficiency is a useful and well-studied notion to classify CRNs. With \Cref{thm:main_result} in place we consider at the end of \Cref{sec:main_result} (and similar as done in \cite{Anderson/CRN/Domination}) its consequences for the case of CRNs with deficiency one (cf.\ \Cref{cor:recurr_defic_one}).}

While formulated in terms of CRNs, the results in this paper equally apply to Petri nets, which is a very well studied model of parallel computation, see, e.g., \cite{DBLP:conf/ac/1996petri1}. Using the ``dictionary'' provided for the reader with a Petri net background (see Subsection~\ref{ssec:crn_petri}), it is straightforward to reformulate the results in this paper in terms of Petri nets.

\ifbool{CONFVER}{Due to space constraints, proofs of the results are omitted and a corollary concerning CRNs of deficiency one is omitted. They can be found in the full version of this paper \cite{B/CRNDom/15}.}{}

\section{Standard graph and CRN/Petri net notions} \label{sec:standard_notions}

\subsection{Preliminaries} \label{ssec:prelim}

Let $\mathbb{N} = \{0,1,\ldots\}$. Let $X$ and $Y$ be arbitrary sets. The set of vectors indexed by $X$ with entries in $Y$ (i.e., the set of functions $\varphi: X \rightarrow Y$) is denoted by $Y^X$. For $v, w \in \mathbb{N}^X$, we write $v \leq w$ if $v(x) \leq w(x)$ for all $x \in X$. Moreover, we write $v < w$ if $v \leq w$ and $v \neq w$. The \emph{support} of $v$, denoted by $\supp(v)$, is the set $\{ x \in X \mid v(x) > 0 \}$. For finite sets $X$ and $Y$, a $X \times Y$ matrix $A$ is a matrix where the rows and columns are indexed by $X$ and $Y$, respectively.

We consider digraphs $G = (V,E,F)$ where $V$ and $E$ are finite sets of vertices and edges and $F: E \rightarrow V^2$ assigns to each edge $e \in E$ an ordered vertex pair $(u,v)$. We denote $V$ by $V(G)$ and $E$ by $E(G)$. The \emph{incidence matrix} of $G$ is the $V(G) \times E(G)$ matrix $A$ where for $e \in E$ with $F(e) = (v,w)$ we have entries $A(v,e) = -1$, $A(w,e) = 1$, and $A(u,e) = 0$ for all $u \in V \setminus \{v,w\}$ if $v \neq w$, and $A(u,e) = 0$ for all $u \in V$ if $v = w$. The number of connected components of a digraph $G$ is denoted by $c(G)$. It is well known that the rank $r(A)$ of the incidence matrix $A$ of a digraph $G$ is equal to $|V|-c(G)$ (where it does not matter over which field the rank is computed \cite[Proposition~5.1.2]{Oxley/MatroidBook-2nd}). From now on we let the field $\mathbb{Q}$ of rational numbers be the field in which we compute.

A walk $\pi$ in $G$ is described by (particular) strings over $E$. Let $\Phi(\pi)$ denote the \emph{Parikh image} of $\pi$, i.e., $\Phi(\pi) \in \mathbb{N}^E$ where $(\Phi(\tau))(e)$ is the number of occurrences of $e$ in $\pi$. We write $\supp(\pi) = \supp(\Phi(\pi))$, i.e., $\supp(\pi)$ is the set of elements that occur in $\pi$. The vectors $v$ of $\ker(A) \cap \mathbb{N}^E$ describe the cycles of $G$, i.e., they describe the Parikh images of closed walks in $G$.

For convenience we identify a digraph $G$ with its $V(G) \times E(G)$ incidence matrix. Hence, we may for example speak of the rank $r(G)$ of $G$. We say that $e \in E(G)$ is a \emph{bridge} if $e$ is not contained in any closed walk of $G$. The \emph{induced subgraph} $G'$ of $G$ with respect to $X \subseteq V(G)$ is the digraph $G' = (X,E',F')$ where $E'$ is the preimage of $X^2$ under $F$ and $F'$ is the restriction of $F$ to $E'$. A \emph{strongly connected component} (\emph{SCC}, for short) is an induced subgraph $G'$ of $G$ with respect to $X \subseteq V(G)$ such that $G'$ contains no bridge and $X$ is largest (with respect to inclusion) with this property.

\subsection{CRNs and Petri nets} \label{ssec:crn_petri}

We now recall the notion of a chemical reaction network.
\begin{definition}
A \emph{chemical reaction network} (or \emph{CRN} for short) $N$ is a $3$-tuple $(S,R,F)$ where $S$ and $R$ are finite sets and $F$ is a function that assigns to each $r \in R$ an ordered pair $F(r) = (v,w)$ where $v,w \in \mathbb{N}^S$. Vector $v$ is denoted by $\tin(r)$ and $w$ by $\tout(r)$.
\end{definition}
The elements of $S$ are called the \emph{species} of $N$, the elements of $R$ are called the \emph{reactions} of $N$, and $F$ is called the \emph{reaction function}. For a reaction $r$, $\tin(r)$ and $\tout(r)$ are called the \emph{reactant vector} and \emph{product vector} of $r$, respectively.

It is common in the literature of CRNs to omit the function $F$ and have $R$ as a set of tuples $(v,w)$. However, this would not allow two different reactions to have the same reactant and product vectors (such situations are common in Petri net theory).

In CRN theory, it is common to write vectors in additive notation, so, e.g., if $S=\{A,B,C\}$, then $A+2B$ denotes the vector $v$ with $v(A) = 1$, $v(B) = 2$, and $v(C)=0$.

\begin{example} \label{ex:toy}
Consider the CRN $N = (S,R,F)$ with $S = \{A,B\}$, $R = \{a,b\}$, $F(a) = (A+B,2B)$ and $F(b) = (B,A)$. This CRN is taken from \cite{Shinar12032010} (see also \cite{Anderson/CRN/Domination}). This example is the running example of this section.
\end{example}

\begin{figure}
\begin{center}
$
\xymatrix{
A+B \ar[r]^{a} & 2B & B \ar[r]^{b} & A
}
$
\end{center}
\caption{The reaction graph of the CRN of \Cref{ex:toy}.}
\label{fig:ex_toy}
\end{figure}
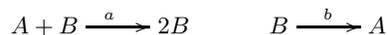

We now define a natural digraph for a CRN $N$, called the reaction graph of $N$. The name is from \cite{DBLP:journals/fuin/MairesseN10}, and the concept is originally defined in \cite{Feinberg/deficiency}.
\begin{definition}
Let $N = (S,R,F)$ be a CRN. The \emph{reaction graph} of $N$, denoted by $\tgr_N$, is the labeled digraph $(V,R,F)$ with $V = \{ \tin(r) \mid r \in R\} \cup \{ \tout(r) \mid r \in R\}$.
\end{definition}
Note that in the reaction graph each reactant and product vector becomes a single vertex. The vertices of the reaction graph are called \emph{complexes}. The reaction graph of the CRN $N$ of our running example (\Cref{ex:toy}) is depicted in \Cref{fig:ex_toy}.

A \emph{configuration} $c$ of $N$ is a vector $c \in \mathbb{N}^S$. Let $r \in R$. We say that $r$ can \emph{fire} on $c$ if $\tin(r) \leq c$. In this case we also write $c \to^r c'$ where $c' = c - \tin(r) + \tout(r)$. Note that $c'$ is a configuration as well. Moreover, we write $c \to c'$ if $c \to^r c'$ for some $r \in R$. For $\tau \in R^*$ (as usual, $R^*$ is Kleene star on $R$) we write $c \to^\tau c'$ if $c \to^{\tau_1} c_1 \cdots \to^{\tau_n} c'$ where $\tau = \tau_1 \cdots \tau_n$ and $\tau_i \in R$ for all $i \in \{1,\ldots,n\}$. The reflexive and transitive closure of the relation $\to$ is denoted by $\to^*$. If $c \to^* c'$, then we say that $c'$ is \emph{reachable} from $c$. We say that a configuration $c$ is \emph{recurrent} if for all $c'$ with $c \to^* c'$ we have $c' \to^* c$. Note that if $c$ is recurrent and $c \to^* c'$, then $c'$ is recurrent.

\begin{example}
Consider again the running example. We have, e.g., $2A+B \rightarrow^{aabb} 2A+B$. However, $2A+B$ is not recurrent as $2A+B \rightarrow^{b} 3A$ and in configuration $3A$ no reaction can fire. In fact, the recurrent configurations of $N$ are precisely those that do not contain any $B$. Indeed, assume $c$ is recurrent. Then we can fire $b$ until we obtain a configuration $c'$ that does not contain any $B$. No reaction can fire for $c'$ and so $c = c'$ since $c$ is recurrent.
\end{example}

The definition of a CRN is equivalent to that of a Petri net \cite{DBLP:conf/ac/1996petri1}. In a Petri net, species are called \emph{places} $p$, reactions are called \emph{transitions}, and configurations are called \emph{markings}. A Petri net is often depicted as a graph with two types of vertices, one type for the places and one for the transitions.
The Petri net-style depiction of the running example is given in \Cref{fig:petri}. The round vertices are the places and the rectangular vertices are the transitions.
We use in this paper several standard Petri net notions, which are recalled in the next subsection.

\tikzstyle{place}=[circle,draw=blue!50,fill=blue!20,thick,
inner sep=0pt,minimum size=6mm]
\tikzstyle{transition}=[rectangle,draw=black!50,fill=black!20,thick,
inner sep=0pt,minimum size=4mm]

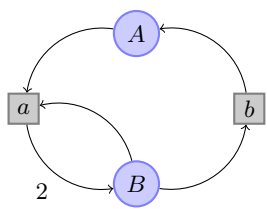
\begin{figure}
\begin{center}
\begin{tikzpicture}[auto]
\node[place] (xA) {$A$};
\node (xd) [below of=xA] {};
\node[place] (xB) [below of=xd] {$B$};
\node[transition] (a) [left of=xd,xshift=-5mm] {$a$};
\node[transition] (b) [right of=xd,xshift=5mm] {$b$};
\draw [->] (xB) to [bend right=45] (a);
\draw [->] (xA) to [bend right=45] (a);
\draw [->] (a) to [bend right=45] node [swap] {2}  (xB);
\draw [->] (xB) to [bend right=45] (b);
\draw [->] (b) to [bend right=45] (xA);
\end{tikzpicture}
\end{center}
\caption{The Petri net-style depiction of the running example.}
\label{fig:petri}
\end{figure}

\subsection{P/T-invariants}

The notions of this subsection are all taken from Petri net theory \cite{DBLP:conf/ac/1996petri1}. We first recall the notion of an incidence matrix of a CRN, which is not to be confused with the notion of an incidence matrix of a digraph (as recalled above). In fact, we will compare in the next subsection the incidence matrix of a CRN with the incidence matrix of its reaction graph.

\begin{definition}\label{def:inci_matrix}
For a CRN $N = (S,R,F)$, the \emph{incidence matrix} of $N$, denoted by $\mathcal{I}_N$, is the $S \times R$ matrix $A$ where for each $r \in R$ the column of $A$ belonging to $r$ is equal to $\tout(r) - \tin(r)$.
\end{definition}

\begin{example}
Consider again the CRN $N$ of the running example. Then
$$
\mathcal{I}_N =
\bordermatrix{ & a & b \cr
A & -1 & 1 \cr
B & 1 & -1}.
$$
\end{example}

Note that if $c \rightarrow^\tau c'$, then $c' = c + \mathcal{I}_N \Phi(\tau)$, where $\Phi(\tau)$ denotes again the Parikh image of $\tau$.

A $v \in \mathbb{N}^S$ is called a \emph{P-invariant} of $N$ if $v^T \mathcal{I}_N = 0$ (here $0$ denotes a zero vector of suitable dimension indexed by $R$). Similarly, $v \in \mathbb{N}^R$ is called a \emph{T-invariant} of $N$ if $\mathcal{I}_N v = 0$, i.e., $v \in \ker(\mathcal{I}_N)$.\footnote{The P and T in P/T-invariant are short for Place and Transition (from Petri net theory). We choose to use these well-known names instead of calling them ``S-invariant'' and ``R-invariant'' for Species and Reaction, respectively.} A P-invariant or T-invariant are also sometimes called P-semiflow and T-semiflow, respectively, in the literature. Observe that if $c \rightarrow^\tau c'$, then $\Phi(\tau)$ is a T-invariant if and only if $c' = c$. A CRN $N$ is called \emph{conservative} if there is a P-invariant $v$ such that $\supp(v) = S$. Also, $N$ is called \emph{consistent} if there is a T-invariant $v$ such that $\supp(v) = R$.

A CRN $N$ is said to be \emph{structurally bounded} when for every configuration $c$, there is a $k_c \in \mathbb{N}$ such that for each configuration $c'$ with $c \to^* c'$ we have that each entry of $c'$ is at most $k_c$. Note that for a structurally-bounded CRN, the number of different configurations reachable from a given configuration is finite, and so for each configuration $c$, there is a recurrent configuration reachable from $c$. In this way, one often informally views the recurrent configurations as the possible states of the CRN in ``the long term''.

The following result is well known\ifbool{CONFVER}{}{, but for completeness we recall its short proof}.
%\begin{restatable}[\cite{DBLP:conf/ac/MemmiR75}]{proposition}{ConsStructB}
\begin{proposition}[\cite{DBLP:conf/ac/MemmiR75}]
\label{prop:cons_structb}
Let $N$ be a CRN. If $N$ is conservative, then $N$ is structurally bounded.
\end{proposition}
%\end{restatable}
\ifbool{CONFVER}{}{
\begin{Proof}
Let $v \in \mathbb{N}^S$ be a P-invariant with $\supp(v) = S$ and let $c$ be a configuration. Let $c \rightarrow^\tau c'$ for some $\tau \in R^*$. We have $c' = c + \mathcal{I}_N \Phi(\tau)$. Thus $v^T c' = v^T c + v^T \mathcal{I}_N \Phi(\tau) = v^T c$ and so for all $s \in S$, $v(s)c'(s) \leq v^Tc$ and therefore $c'(s) \leq v^T c / v(s)$.
\end{Proof}
}

\begin{example}
The CRN $N$ of the running example is both conservative and consistent. Indeed, any $v \in \mathbb{N}^S$ with $v(A)=v(B) \geq 1$ is a P-invariant with $\supp(v) = S$ and any $w \in \mathbb{N}^R$ with $v(a)=v(b) \geq 1$ is T-invariant with $\supp(v) = R$.
\end{example}

\subsection{Deficiency} \label{ssec:defic}
The notions that we recall in this subsection are originally from chemical reaction theory (and are less studied within Petri net theory).

Let $N  = (S,R,F)$ be a CRN and let $V = \{ \tin(r) \mid r \in R\} \cup \{ \tout(r) \mid r \in R\}$. We denote by $\mathcal{Y}_N$ the $S \times V$ matrix with for all $s \in S$ and $v \in V$, entry $\mathcal{Y}_N(s,v)$ is equal to $v(s)$.

The next lemma relates the incidence matrix $\mathcal{I}_N$ of a CRN $N$ with the incidence matrix of the reaction graph $\tgr_N$ of $N$.
%\begin{restatable}[Section~6 of \cite{FeinbergHorn/deficiency}]{lemma}{RGvsIM}
\begin{lemma}[Section~6 of \cite{FeinbergHorn/deficiency}]
\label{lem:deficiency_matrix}
Let $N = (S,R,F)$ be a CRN. Then $\mathcal{I}_N = \mathcal{Y}_N \tgr_N$.
\end{lemma}
%\end{restatable}
In the above equality, $\tgr_N$ denotes the incidence matrix $\tgr_N$ and not the graph.
\ifbool{CONFVER}{}{
\begin{Proof}
Let $V = \{ \tin(r) \mid r \in R\} \cup \{ \tout(r) \mid r \in R\}$. Let $p \in P$ and $r \in R$. Then $\mathcal{I}_N(p,r) = (\tout(r) - \tin(r))(p) = \mathcal{Y}_N(p,\tout(r)) \cdot 1 + \mathcal{Y}_N(p,\tin(r)) \cdot (-1) = \sum_{x \in V} \mathcal{Y}_N(p,x) \tgr_N(x,r) \allowbreak = \mathcal{Y}_N \tgr_N$.
\end{Proof}
}

As a corollary to \Cref{lem:deficiency_matrix}, we have the following.
\begin{corollary}[\cite{DBLP:journals/fuin/MairesseN10}] \label{cor:deficiency}
Let $N = (S,R,F)$ be a CRN. Then $\ker(\tgr_N) \subseteq \ker(\mathcal{I}_N)$.
\end{corollary}

The vectors $v$ of $\ker(\tgr_N) \cap \mathbb{N}^R$, which are T-invariants by \Cref{cor:deficiency}, are called \emph{closed} T-invariants \cite{Boucherie/ClosedTInvariant}. Recall that the vectors $v$ of $\ker(\tgr_N) \cap \mathbb{N}^R$ describe the cycles of $\tgr_N$, and so for each closed T-invariant $v$ of $N$, $\supp(v)$ does not contain any bridge of $\tgr_N$. Since each of the entries of a T-invariant is nonnegative, the linear space $\ker(\mathcal{I}_N)$ does not necessarily have a basis consisting of only T-invariants, see \Cref{ex:defic} below.

The \emph{deficiency} $\delta(N)$ of a CRN $N$ is $r(\tgr_N) - r(\mathcal{I}_N)$. By \Cref{cor:deficiency}, $\delta(N)$ is non-negative. Thus, one may view $\delta(N)$ as a measure of the difference in dimensions between $\ker(\tgr_N)$ and $\ker(\mathcal{I}_N)$. The former is determined only by the structure of the reaction graph (ignoring the identity of the vertices), while the latter also incorporates the relations that rely on the identities of the vertices of the reaction graph.

Recall from Subsection~\ref{ssec:prelim} that $r(\tgr_N) = |V(\tgr_N)| - c(\tgr_N)$. Hence, we have $\delta(N) = |V(\tgr_N)| - c(\tgr_N) - r(\mathcal{I}_N)$ \cite{Horn/deficiency,Feinberg/deficiency}. Note that if $\delta(N) = 0$, then every T-invariant of $N$ is closed and $\ker(\tgr_N) = \ker(\mathcal{I}_N)$.

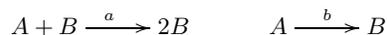
\begin{figure}
\begin{center}
$
\xymatrix{
A+B \ar[r]^{a} & 2B & A \ar[r]^{b} & B
}
$
\end{center}
\caption{The reaction graph of a CRN discussed in \Cref{ex:defic}.}
\label{fig:ex_toy_inv}
\end{figure}

\begin{example} \label{ex:defic}
In the running example, $\ker(\tgr_N)$ only contains the zero vector, while $\ker(\mathcal{I}_N)$ contains all scalar multiples of the vector $w$ with $w(a)=w(b)=1$. Thus $\ker(\mathcal{I}_N)$ has a basis consisting of only T-invariants. Moreover, $\delta(N) = 1$. Alternatively, the reaction graph $\tgr_N$ has $4$ vertices and $2$ connected components and $r(\mathcal{I}_N) = 1$. Thus, $\delta(N) = 4-2-1 = 1$.

If we consider the CRN $N'$ of \Cref{fig:ex_toy_inv}, then $\ker(\tgr_{N'})$ also only contains the zero vector, while $\ker(\mathcal{I}_{N'})$ contains all scalar multiples of the vector $w$ with $w(a)=-w(b)=1$. Again, $\delta(N') = 1$, however the only T-invariant of $\ker(\mathcal{I}_{N'})$ is the zero vector.
\end{example}

\section{Dominance and non-closed T-invariants} \label{sec:main_result}

Note that there is a natural partial order for the set of SCCs of a graph: for SCCs $X$ and $Y$, we have $X \preccurlyeq Y$ if there is a path from a vertex of $Y$ to a vertex of $X$. We now consider a different partial order, denoted by $\leq_d$, for the SCCs of a reaction graph of a CRN.

Let $N$ be a CRN. For SCCs $X$ and $Y$ of $\tgr_N$ we write $X \leq_d Y$ if there are vertices $x$ of $X$ and $y$ of $Y$ such that $x \leq y$.

%\begin{restatable}{lemma}{SCCpo}
\begin{lemma}
\label{lem:SCC_po}
Let $N = (S,R,F)$ be a structurally-bounded CRN. Then the $\leq_d$ relation between SCCs of $\tgr_N$ is a partial order.
\end{lemma}
%\end{restatable}
\ifbool{CONFVER}{}{
\begin{Proof}
The $\leq_d$ relation is obviously reflexive and transitive. To show that $\leq_d$ is antisymmetric, let $X \leq_d Y$ and $Y \leq_d X$ for some SCCs $X$ and $Y$ of $\tgr_N$. Hence there are vertices $x_1$ and $x_2$ of $X$ and $y_1$ and $y_2$ of $Y$ such that $x_1 \leq y_1$ and $y_2 \leq x_2$. Let $\pi_1$ be a path from $x_1$ to $x_2$ and let $\pi_2$ be a path from $y_2$ to $y_1$ in $\tgr_N$. Then $x_1+y_2 \to^{\pi_1} x_2+y_2 \to^{\pi_2} x_2+y_1$. If $X \neq Y$, then $x_1 < y_1$ and $y_2 < x_2$. Thus we have $x_1+y_2 < x_2+y_1$, and so $N$ is not structurally bounded --- a contradiction.
\end{Proof}
}

For SCCs $X$ and $Y$ we write $X <_d Y$ if $X \leq_d Y$ and $X \neq Y$. We say that $X$ \emph{dominates} $Y$ when $X <_d Y$. For a set $\mathcal{S}$ of SCCs, we let $\min_{\leq_d}(\mathcal{S}) \subseteq \mathcal{S}$ be the set of elements of $\mathcal{S}$ that are minimal with respect to the $\leq_d$ relation among all the elements of $\mathcal{S}$.

Let us define for a SCC $X$ of $\tgr_N$, $\CCout(X) = \{ r \in E(\tgr_N) \mid \tin(r) \in V(X), \tout(r) \notin V(X) \}$. We call $X$ \emph{terminal} if $\CCout(X) = \emptyset$. We call a reaction $r$ (complex $x$, resp.) \emph{terminal} if $r \in E(X)$ ($x \in V(X)$, resp.) for some terminal SCC $X$ of $\tgr_N$.

We will consider the minimal set $\mathcal{X}$ of non-terminal SCCs that dominates all other non-terminal SCCs. In other words, if we let $\mathcal{N}$ be the set of non-terminal SCCs, then $\mathcal{X} = \min_{\leq_d}(\mathcal{N})$.

Let $B$ be the set of bridges of $\tgr_N$. The \emph{exit set} of a set $\mathcal{S}$ of non-terminal SCCs, is a set $Z \subseteq B$ with both $|Z| = |\mathcal{S}|$ and $|Z \cap \CCout(X)| = 1$ for all $X \in \mathcal{S}$. In other words, $Z$ contains exactly one bridge of $\CCout(X)$ for each $X \in \mathcal{S}$.

Assuming the existence of a non-terminal reaction that can fire for some recurrent configuration $c$, the main result of this paper ensures the existence of certain sequences $\tau$ with $c' \rightarrow^{\tau} c'$ for some configuration $c'$ reachable from $c$. For each exit set $Z$, there exists such a $\tau$ that avoids all bridges outside $Z$ and, at the same time, uses the bridges of $Z$ whenever possible. As a consequence, each of the sequences $\tau$ corresponds to a T-invariant $v = \Phi(\tau)$ that have zero entries for the bridges outside $Z$ and nonzero entries for some of the bridges inside $Z$. We will show that for various CRNs this necessary condition allows one to show that only terminal reactions can fire for all its recurrent configurations.

The proof idea is the following. Let us start with a recurrent configuration $c$.  While traversing the configuration space by applying reactions starting from $c$, we need never choose a bridge of $\tgr_N$ going out of a SCC $X$ that is dominated by some $Y$ (i.e., $Y <_d X$). Indeed, if $x \in V(X)$ and $y \in V(Y)$ with $y < x$, then we may walk inside $X$ to $x$ and $y < x$ implies that any reaction $r$ with $\tin(r) = y$ can fire for $x$. In this way we also avoid taking a reaction $r'$ with $\tin(r') = x$. Moreover, walking out of $Y$ can be done by taking any of the bridges. We choose the one from the exit set $Z$. Now, eventually, our path inside $\tgr_N$ will lead to a terminal vertex. However, since $c$ is recurrent, we can go back to $c$. If a non-terminal reaction can fire for $c$, then this means that we can iterate this process (walking along bridges, etc.). Structural boundedness finally ensures that the configuration space is finite and so, we must eventually repeat a configuration that closes the ``circuit''.

We are now ready to formulate the main result of this paper.
%\begin{restatable}{theorem}{MainResult}
\begin{theorem}
\label{thm:main_result}
Let $N = (S,R,F)$ be a structurally-bounded CRN, and let $\mathcal{X} = \min_{\leq_d}(\mathcal{N})$, where $\mathcal{N}$ is the set of non-terminal SCCs of $\tgr_N$. Let $B$ be the set of bridges of $\tgr_N$. Let $L$ be the set of all non-terminal reactions $r$ of $\tgr_N$ such that there is a non-terminal reaction $r'$ of $\tgr_N$ with $\tin(r') < \tin(r)$.

If some non-terminal reaction can fire for some recurrent configuration $c$, then for each exit set $Z$ of $\mathcal{X}$, there is a $\tau \in R^*$ such that
%
%(1) $\supp(\tau) \cap B \subseteq Z$,
%(2) $\supp(\tau) \cap L = \emptyset$,
%(3) $\tau = \tau_1 \cdots \tau_n$ where each $\tau_i$ is a path in $\tgr_N$ from a non-terminal vertex to a terminal vertex, and
%(4) $c' \rightarrow^{\tau} c'$ for some recurrent configuration $c'$ reachable from $c$.
%
\begin{enumerate}
\item $\tau$ contains no reactions from $(B \setminus Z) \cup L$,
\item $\tau = \pi_1 \sigma_1 \cdots \pi_n \sigma_n$ where each $\pi_i$ is a path in $\tgr_N$ from a non-terminal vertex to a terminal vertex and each $\sigma_i$ is a sequence of terminal reactions, and \label{item:main_thm_seq_paths}
\item $c' \rightarrow^{\tau} c'$ for some recurrent configuration $c'$ reachable from $c$.
\end{enumerate}
\end{theorem}
%\end{restatable}
\ifbool{CONFVER}{}{
\begin{Proof}
Assume that some non-terminal reaction $r_1$ can fire for some recurrent configuration $c$. Let $Z \subseteq B$ be an exit set of $\mathcal{X}$.

Let $Y$ be the SCC containing vertex $\tin(r_1)$ and let $s_1 \in R^*$ be a shortest path in $Y$ from $\tin(r_1)$ to some vertex $y$ with $x \leq y$ for some vertex $x$ of some $X \in \mathcal{X}$. Observe that $\supp(s_1) \cap L = \emptyset$ (if $Y \in \mathcal{X}$ or $r_1 \in L$, then $s_1$ is the empty string). Let $s_2 b_X \in R^*$ be a path in $\tgr_N$ such that $s_2$ is a path in $X$ from $x$ and $b_X \in Z \cap \CCout(X)$. Then $\alpha_1 = r_1 s_1 s_2 b_X$ can fire for $c$. We repeat this procedure for vertex $\tout(b_X)$. In this way, we obtain a path $\alpha_1 \alpha_2$. By iteration, we obtain a path $\pi_1 = \alpha_1 \cdots \alpha_l$ from $\tin(r_1)$ to a terminal vertex $x$. Note that $\pi_1$ can fire for $c$. Since $c$ is recurrent, there is a sequence $\sigma_1$ of terminal reactions such that $\tau_1 = \pi_1 \sigma_1$ has the property that $c \to^{\tau_1} c'$ where some non-terminal reaction $r_2$ can fire for $c'$. Note that both $\tau_1$ contains no reactions from $(B \setminus Z) \cup L$. Also, $\pi_1$ is a path in $\tgr_N$ from a non-terminal vertex to a terminal vertex.

We repeat the above described procedure (that constructed $\tau_1$) for configuration $c'$ and vertex $\tin(r_2)$, to obtain (by iteration) an infinite sequence $\tau = \tau_1 \tau_2 \cdots$. Since $\tau$ is infinite and $N$ is structurally bounded, there is a configuration $c_r$ such that $c \to^{\tau_{\mathrm{pre}}} c_r \to^{\tau_{\mathrm{loop}}} c_r$ and $\tau_{\mathrm{loop}} = \tau_i \cdots \tau_j$ for some $i<j$. Note that, by the construction of $\tau$, $\tau_{\mathrm{loop}}$ contains no reactions from $(B \setminus Z) \cup L$, and so we are done.
\end{Proof}
}

\begin{figure}
\begin{center}
$
\xymatrix{
A \ar[r]^{a} & E & C \ar[r]^b & D & E + D \ar[r]^{d} & A + C
}
$
\end{center}
\caption{The reaction graph of the CRN of \Cref{ex:recurr}.}
\label{fig:ex_recurr}
\end{figure}
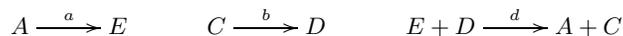

We illustrate \Cref{thm:main_result} through a couple of examples.
\begin{example}\label{ex:recurr}
Consider the CRN $N$ of \Cref{fig:ex_recurr}. It is easy to verify that $c=A+C$ is a recurrent configuration. Moreover, there is a non-terminal reaction $r$ that can fire for this configuration (take $r=a$ or $r=b$). Note that there is only one exit set $Z$ for $\mathcal{X}$, which is $Z = B = \{a,b,d\}$. By Theorem~\ref{thm:main_result}, there is a $\tau \in R^*$ such that (1) $\tau$ contains no reactions from $(B \setminus Z) \cup L$, (2) $\tau$ is a sequence of paths, each going to a terminal vertex, and (3) $c' \rightarrow^{\tau} c'$ for some recurrent configuration $c'$ reachable from $c$. Indeed, we can choose, e.g., $\tau = abd$ and $c' = A+C$.
\end{example}

\begin{figure}
\begin{center}
$
\xymatrix{
A + C \ar[r]^{a} & E + C & E + D \ar[r]^b & A + D
}
$
\end{center}
\caption{The reaction graph of the CRN of \Cref{ex:recurr2}.}
\label{fig:ex_recurr2}
\end{figure}
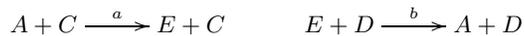

We now give another example.
\begin{example}\label{ex:recurr2}
Consider the CRN $N$ of \Cref{fig:ex_recurr2}. It is easy to verify that $c=A+C+D$ is a recurrent configuration. Moreover, there is a non-terminal reaction $r$ that can fire for this configuration (take $r=a$). We have that $Z = B = \{a,b\}$ is the unique exit set $Z$ for $\mathcal{X}$. We notice that $\tau = ab$ and $c'=A+C+D$ satisfy the conditions of Theorem~\ref{thm:main_result}. Indeed, we have $\tau = \tau_1 \sigma_1 \tau_2 \sigma_2$ with $\tau_1 = a$, $\sigma_1 = \epsilon$ (the empty string), $\tau_2 = b$, and $\sigma_2 = \epsilon$. Note that if $N$ contained the additional reaction $A + D \to^d E + D$, then $\tau = ab$ and $c'=A+C+D$ would again satisfy the conditions of Theorem~\ref{thm:main_result}, where $\tau = \tau_1 \sigma_1$ with $\tau_1 = a$ and $\sigma_1 = b$.
\end{example}
% Another example is $A+C \to B+C$ and $B+D \to A+D$.

Considering the non-closed T-invariant $v = \Phi(\tau)$ with $\tau$ from \Cref{thm:main_result}, we have the following corollary to \Cref{thm:main_result}. Note that Condition~\ref{item:main_thm_seq_paths} of \Cref{thm:main_result} implies that $\supp(v)$ contains a bridge, and therefore $v(z) \neq 0$ for some $z \in Z$.
\begin{corollary}
\label{cor:main_result}
Let $N$, $\mathcal{X}$, $B$, and $L$ be as in \Cref{thm:main_result}.

Assume there is an exit set $Z$ of $\mathcal{X}$ such that there is no non-closed T-invariant $v$ with (1) $v(x) = 0$ for all $x \in (B \setminus Z) \cup L$ and (2) $v(z) \neq 0$ for some $z \in Z$.
%$\emptyset \neq \supp(v) \cap B \subseteq Z$ and $\supp(v) \cap L = \emptyset$.

Then no non-terminal reaction can fire for any recurrent configuration of $N$.
\end{corollary}

We remark that, in view of \Cref{thm:main_result}, \Cref{cor:main_result} can be strengthened by replacing the condition $v(z) \neq 0$ for some $z \in Z$
%$\emptyset \neq \supp(v) \cap B$
with the stronger (but more involved) condition that says that the (occurrences of the) non-terminal reactions of $v$ form a set of paths where each path ends in a terminal vertex.

Note that since closed T-invariants $v$ cannot contain bridges, we may without loss of generality remove the condition that $v$ is  ``non-closed'' in \Cref{cor:main_result}.

We use \Cref{cor:main_result} to determine whether no non-terminal reaction can fire for any recurrent configuration of a CRN. While non-closed T-invariants have a central role in \Cref{cor:main_result}, curiously, this notion from \cite{Boucherie/ClosedTInvariant} has been given only modest attention in both the Petri net theory and the CRN theory.

For a given exit set $Z$ of $\mathcal{X}$, one can verify using linear programming in polynomial time whether or not there is a non-closed T-invariant $v$ with the properties of \Cref{cor:main_result}. While in general there may be an exponential number of exit sets (exponential in the number of reactions) to check, in many cases the number of exit sets is severely constraint and in these cases the sufficient condition of \Cref{cor:main_result} is computationally efficient.

%\corr{TODO: definitie $V$?}
%Indeed, subject to $\mathcal{I}_N v = 0$ and $v \geq 0$ one maximizes $w^T v$, where $w(x) = 1$ for all $x \in Z$, $w(x) = -\inf$ for all $x \in (B \setminus Z) \cup L$, and $w(x) = 0$ for all $x \in V \setminus B$. Any nonzero solution $v$ corresponds to a closed T-invariant by multiplying by its greatest common divisor.

%This  since determining $\ker(\mathcal{I}_N)$ (which in turn determines the T-invariants) can be done using Gaussian elimination which has complexity $O(n^3)$ \corr{eigenlijk linear programming?}, where $n$ is the number of rows/columns (the largest of the two) of $\mathcal{I}_N$.

\begin{figure}
\begin{center}
$
\xymatrix{
A \ar@/^5pt/[r]^{a} & J \ar@/^5pt/[l]^{b} \ar@/^5pt/[r]^{c} & C \ar@/^5pt/[l]^{d} \ar[r]^{e} & D & D+E \ar[r]^{f} & J+H\\
& A+H \ar[r]^{g} & A+E & & C+H \ar[r]^{h} & C+E\\
}
$
\end{center}
\caption{The reaction graph of the CRN of \Cref{ex:from_lit}.}
\label{fig:ex_from_lit}
\end{figure}
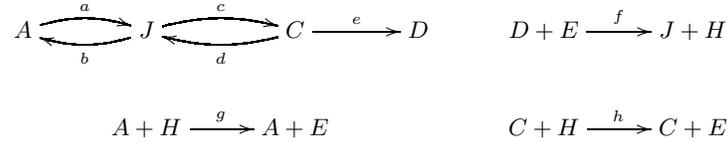

We now give some examples to illustrate \Cref{cor:main_result}.
\begin{example} \label{ex:from_lit}
Consider the CRN $N$ of \Cref{fig:ex_from_lit}. This CRN is a simplification of a CRN from biology studied in \cite{Shinar12032010} (see also \cite{Anderson/CRN/Domination}).
We have
$$
\mathcal{I}_N =
\bordermatrix{
  & a  & b  & c  & d  & e  & f  & g  & h \cr
A & -1 & 1  & 0  & 0  & 0  & 0  & 0  & 0 \cr
J & 1  & -1 & -1 & 1  & 0  & 1  & 0  & 0 \cr
C & 0  & 0  & 1  & -1 & -1 & 0  & 0  & 0 \cr
D & 0  & 0  & 0  & 0  & 1  & -1 & 0  & 0 \cr
E & 0  & 0  & 0  & 0  & 0  & -1 & 1  & 1 \cr
H & 0  & 0  & 0  & 0  & 0  & 1  & -1 & -1
}.
$$
It is easy to verify that the sum of the rows of $\mathcal{I}_N$ is the zero vector and so $N$ is conservative. Consequently, $N$ is structurally bounded. It turns out that $\ker(\mathcal{I}_N)$ is of dimension $4$ and is spanned by T-invariants. In fact, one can verify that $\ker(\mathcal{I}_N)$ is spanned by the two closed T-invariants $w_1 = \Phi(ab)$ and $w_2 = \Phi(cd)$ together with the two non-closed T-invariants $v_1 = \Phi(gfce)$ and $v_2 = \Phi(hfce)$. We remark that $A+H+D \to^{gfce} A+H+D$ and $C+H+D \to^{hfce} C+H+D$. Thus $\delta(N) = 2$. Note that $B = \{e,f,g,h\}$ is the set of bridges of $\tgr_N$. Let $\mathcal{X}$ be the set of non-terminal SCCs of $\tgr_N$ that are minimal with respect to $\leq_d$. We notice that $Z = \{e,f\}$ is the only exit set of $\mathcal{X}$. Also $L = \{g,h\}$. Now, the non-closed T-invariants $v_1$ and $v_2$ are witnesses that there is no non-closed T-invariant $v$ with both
(1) $v(g) = v(h) = 0$ (note that $(B \setminus Z) \cup L = \{g,h\}$) and (2) either $v(e)$ or $v(f)$ nonzero. By \Cref{cor:main_result}, for every recurrent configuration no non-terminal reaction can fire. Since every reaction is non-terminal, for every recurrent configuration no reaction can fire.
\end{example}

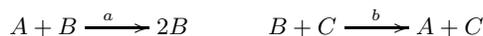
\begin{figure}
\begin{center}
$
\xymatrix{
A+B \ar[r]^{a} & 2B & B+C \ar[r]^{b} & A+C
}
$
\end{center}
\caption{The reaction graph of the CRN of \Cref{ex:no_converse}.}
\label{fig:ex_no_converse}
\end{figure}

The next example shows that the converse of \Cref{thm:main_result} does not hold.
\begin{example} \label{ex:no_converse}
Consider the CRN $N$ of \Cref{fig:ex_no_converse}. We show that no reaction can fire for any recurrent configuration of $N$. Let $c$ be a recurrent configuration. If $c$ does not contain any $C$, then we can fire reaction $a$ until we obtain a configuration $c'$ for which no more reactions can fire. Since $c$ is recurrent, $c = c'$ and we are done. If $c$ contains at least one $C$, then we can apply reaction $b$ until we obtain a configuration $c''$ with only $A$'s and $C$'s. Hence no reaction can fire for $c''$. Since $c$ is recurrent, we have $c = c''$ and we are done.

However, for $c = A+B+C$ we have $c \rightarrow^{\tau} c$ with $\tau = ab$. We notice that $Z = \{a,b\}$ is the only exit set of $\mathcal{X}$ and $(B \setminus Z) \cup L = \emptyset$. Thus $\tau$ trivially contains no reactions from $(B \setminus Z) \cup L$ and $\tau = \pi_1\pi_2$ with paths $\pi_1 = a$ and $\pi_2 = b$ in $\tgr_N$ from non-terminal vertices to terminal vertices. This shows that the converse of \Cref{thm:main_result} does not hold.

We remark that if we remove species $C$ from reaction $b$\ifbool{CONFVER}{ (in this way obtaining the running example of \Cref{sec:standard_notions})}{}, then \Cref{cor:main_result} (and \Cref{thm:main_result}) would have been applicable to show that no (non-terminal) reaction can fire for any recurrent configuration of $N$.
\end{example}

\ifbool{CONFVER}{}{
We now consider the case where the deficiency is $1$. This severely restricts the structure of the non-closed T-invariants.
%\begin{restatable}{lemma}{DeficOneAllBridge}
\begin{lemma}
\label{lem:deficiency_one}
Let $N = (S,R,F)$ be a consistent CRN with $\delta(N)=1$. Then for all non-closed T-invariants $v$, $\supp(v)$ contains every bridge of $\tgr_N$.
\end{lemma}
%\end{restatable}
\ifbool{CONFVER}{}{
\begin{Proof}
Let $v$ be a non-closed T-invariant and let $b$ be a bridge of $\tgr_N$. Since $N$ is consistent, there is a T-invariant $w$ with $\supp(w) = R$. Since $b \in R$, $w$ is non-closed. Thus $v, w \in \ker(\mathcal{I}_N) \setminus \ker(\tgr_N)$. Since $\delta(N)=1$ and $\supp(z)$ cannot contain any bridge for $z \in \ker(\tgr_N)$, we have that $b \in \supp(w)$ if and only if $b \in \supp(v)$. Hence $b \in \supp(v)$.
\end{Proof}
}

The next result is essentially Theorem~3.5 of the supplementary material of \cite{Anderson/CRN/Domination} (although there it is stated in terms of notions from mass-action kinetics and stochastics), and follows directly from \Cref{thm:main_result} and \Cref{lem:deficiency_one}.
%\begin{restatable}[\cite{Anderson/CRN/Domination}]{corollary}{DeficOneRecurr}
\begin{corollary}[\cite{Anderson/CRN/Domination}]
\label{cor:recurr_defic_one}
Let $N$ be a structurally-bounded and consistent CRN with $\delta(N)=1$. If there are non-terminal vertices $x$ and $y$ such that $x < y$, then for all recurrent configurations $c$, none of the non-terminal reactions can fire.
\end{corollary}
%\end{restatable}
\ifbool{CONFVER}{}{
\begin{Proof}
Assume there are non-terminal reactions $x$ and $y$ such that $x < y$ and assume to the contrary that some non-terminal reaction $r$ can fire for some recurrent configuration $c$. By \Cref{lem:deficiency_one}, for all non-closed T-invariants $v$, $\supp(v)$ contains every bridge of $\tgr_N$. Hence, by \Cref{thm:main_result}, every non-terminal SCC of $\tgr_N$ is minimal with respect to $\leq_d$ among the non-terminal SCCs of $\tgr_N$ --- a contradiction by the existence of $x$ and $y$ (note that $x$ and $y$ cannot be vertices of the same SCC since $N$ is structurally bounded).
\end{Proof}
}

\begin{example}
Consider the CRN $N$ of the running example of \Cref{sec:standard_notions}. Recall that $N$ is conservative, and therefore $N$ is structurally bounded. Also recall that $\delta(N)=1$. By \Cref{cor:recurr_defic_one}, no non-terminal reaction can fire for any recurrent configuration $c$ of $N$. Since all reactions of $N$ are non-terminal, no reaction can fire for any recurrent configuration $c$ of $N$.
\end{example}
}

\section{Using rates} \label{sec:using_rates}

This paper is inspired by the main technical result of \cite{Anderson/CRN/Domination} (cf.\ Theorem~3.3 of the supplementary material of \cite{Anderson/CRN/Domination}). In this section we recall its result. First we recall a particular matrix. Let $\mathbb{R}_{\geq 0}$ ($\mathbb{R}_{> 0}$, resp.) be the set of nonnegative (positive, resp.) real numbers.

\begin{definition}
Let $N = (S,R,F)$ be a CRN. Let $V = V(\tgr_N)$ and let $\kappa \in \mathbb{R}_{> 0}^R$. We denote by $\mathcal{K}_{N,\kappa}$ the $S \times V$ matrix where for each $x \in V$ the column of $\mathcal{K}_{N,\kappa}$ belonging to $x$ is equal to $\sum_{r \in R, \tin(r) = x} \kappa(r)\cdot(\tout(r) - \tin(r))$.
\end{definition}
The value $\kappa(r)$ in \Cref{thm:anderson} may be interpreted as the ``rate'' of reaction $r$. Note that the definition of $\mathcal{K}_{N,\kappa}$ is closely related to the definition of $\mathcal{I}_N$ (\Cref{def:inci_matrix}).

We are now ready to formulate the main technical result of \cite{Anderson/CRN/Domination}.
\begin{theorem}[\cite{Anderson/CRN/Domination}] \label{thm:anderson}
Let $N = (S,R,F)$ be a conservative CRN and $V = V(\tgr_N)$. Let $L$ be the set of non-terminal vertices $v$ of $\tgr_N$ such that there is a non-terminal vertex $v'$ of $\tgr_N$ with $v' < v$. Assume that $L \neq \emptyset$.

If some non-terminal reaction can fire for some recurrent configuration $c$, then for all $\kappa \in \mathbb{R}_{> 0}^R$, there is a $w \in \ker(\mathcal{K}_{N,\kappa}) \cap \mathbb{R}_{\geq 0}^V$ with $\supp(w) \cap L = \emptyset$ and there is a non-terminal vertex $x$ with $x \in \supp(w)$.
\end{theorem}

\Cref{thm:anderson} is proved in \cite{Anderson/CRN/Domination} using both intricate probabilistic arguments and methods from mass action kinetics. In \cite{Anderson/CRN/Domination}, the theorem is unnecessarily stated in a probabilistic fashion using the notion of ``positive recurrent configuration'' for stochastically modeled CRNs: it can be stated in a deterministic way (see \Cref{thm:anderson} above) by realizing that the configuration space is finite for a given initial configuration in a structurally-bounded CRN. This deterministic formulation and the discrete model (in contrast to mass action) triggered the search of this paper for a combinatorial explanation of this result. We invite the reader to compare the proof techniques used to prove \Cref{thm:anderson} in \cite{Anderson/CRN/Domination} and \Cref{thm:main_result} in this paper. \ifbool{CONFVER}{}{In \cite{Anderson/CRN/Domination}, \Cref{cor:recurr_defic_one} is proved using \Cref{thm:anderson} while in this paper it is shown using \Cref{thm:main_result}.}

\begin{figure}
\begin{center}
$
\xymatrix{
2B & A+B \ar[r]^{a} \ar[l]_{b} & 2A
}
$
\end{center}
\caption{The reaction graph of the CRN of \Cref{ex:2out}.}
\label{fig:ex_2out}
\end{figure}
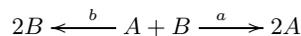

Note that if $L = \emptyset$, then \Cref{thm:anderson} is silent. We now show an example with $L = \emptyset$ where \Cref{cor:main_result} can be applied.
\begin{example} \label{ex:2out}
Consider the CRN $N$ of \Cref{fig:ex_2out}. Note that $N$ is conservative with $w(A) = w(B) = 1$ as a witness. The only T-invariants $v$ of $N$ are those where $v(a) = v(b)$. Let $Z = \{a\}$ be an exit set of $\mathcal{X}$. Then there is no non-closed T-invariant $v$ with $v(b) = 0$ and $v(a) \neq 0$. By \Cref{cor:main_result}, no non-terminal reaction can fire for any recurrent configuration $c$ of $N$. Since all reactions of $N$ are non-terminal, no reaction can fire for any recurrent configuration $c$ of $N$. Indeed, one observes that the recurrent configurations of $N$ are those configurations containing either only $A$'s or only $B$'s, for which $a$ and $b$ cannot fire.
\end{example}
We conjecture that the assumption $L \neq \emptyset$ can be removed from \Cref{thm:anderson}. In case $L \neq \emptyset$ is removed from \Cref{thm:anderson}, then \Cref{thm:anderson} also predicts that no non-terminal reaction can fire for any recurrent configuration of the CRN of \Cref{ex:2out}. Next, we give an example with $L \neq \emptyset$, where \Cref{cor:main_result} can be applied but \Cref{thm:anderson} is silent.

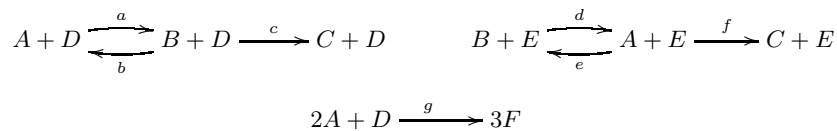
\begin{figure}
\begin{center}
$
\xymatrix@R=6mm{
A+D \ar@/^5pt/[r]^{a} & B+D \ar@/^5pt/[l]^{b} \ar[r]^{c} & C+D & B+E \ar@/^5pt/[r]^{d} & A+E \ar@/^5pt/[l]^{e} \ar[r]^{f} & C+E \\
 & & 2A+D \ar[r]^{g} & 3F
}
$
\end{center}
\caption{The reaction graph of the CRN of \Cref{ex:not_silent}.}
\label{fig:ex_not_silent}
\end{figure}

\begin{example} \label{ex:not_silent}
Consider the CRN $N$ of \Cref{fig:ex_not_silent}. Note that $N$ is conservative with $w(X) = 1$ for all species $X$ as a witness. Note that $A+D < 2A+D$ and so $L \neq \emptyset$ in \Cref{thm:anderson}. Let $\kappa \in \mathbb{R}_{> 0}^R$. We have $\mathcal{K}_{N,\kappa} = $
$$
\resizebox{\textwidth}{!}{
\bordermatrix{
  & A+D  & B+D  & B+E & A+E  & 2A+D & C+D & C+E & 3F \cr
A & -\kappa(a) & \kappa(b)    & \kappa(d)   & -\kappa(e)-\kappa(f) & -2\kappa(g) & 0 & 0 & 0 \cr
B & \kappa(a)  & -\kappa(b)-\kappa(c) & -\kappa(d)  & \kappa(e)    & 0    & 0 & 0 & 0 \cr
C & 0    & \kappa(c)    & 0   & \kappa(f)    & 0    & 0 & 0 & 0 \cr
D & 0    & 0    & 0   & 0    & -\kappa(g)   & 0 & 0 & 0 \cr
E & 0    & 0    & 0   & 0    & 0    & 0 & 0 & 0 \cr
F & 0    & 0    & 0   & 0    & 3\kappa(g)   & 0 & 0 & 0
}.}
$$
Let $w \in \mathbb{R}_{\geq 0}^V$ with $\kappa(a) w(A+D) = \kappa(d) w(B+E) > 0$ and $w(x) = 0$ for all other $x \in V$. Then $w \in \ker(\mathcal{K}_{N,\kappa}) \cap \mathbb{R}_{\geq 0}^V$ with $x \in \supp(w)$ for some non-terminal vertex $x$ and $\supp(w) \cap L = \emptyset$. Thus \Cref{thm:anderson} is silent. On the other hand, none of the non-closed T-invariants of $N$ contains a bridge and so by \Cref{cor:main_result}, no non-terminal reaction can fire for any recurrent configuration of $N$.
\end{example}

Conversely, despite trying numerous examples, we could not find an example where \Cref{thm:anderson} predicts that no non-terminal reaction can fire for any recurrent configuration, but where \Cref{thm:main_result} is silent.

\section{Discussion}
Based on structural properties of CRNs, the main result of this paper (cf.\ \Cref{thm:main_result}) provides a sufficient condition to analyze the long-term behavior of CRNs. While its proof is using basic combinatorial arguments, the result is powerful enough to apply to a large class of CRNs. Also, the sufficient condition is computationally-efficient to verify for many CRNs. Another such sufficient condition is shown in \cite{Anderson/CRN/Domination}, cf.\ \Cref{thm:anderson}. We have shown examples of CRNs where \Cref{thm:main_result} is applicable while \Cref{thm:anderson} is silent.

Given that discrete CRNs are equivalent to Petri nets, it is curious that the corresponding research areas of CRN theory and Petri net theory have evolved almost independently. In this paper we shown that notions from Petri net theory (in particular, T-invariance) are useful for CRN theory. Similarly, notion such as deficiency, originating from CRN theory, are useful for Petri net theory. At the interface of these two notions is the scarcely-studied notion of non-closed T-invariant, which is crucial in the sufficient condition of \Cref{cor:main_result}. This illustrates that both research areas can significantly profit from each other.

An open problem is resolving whether \Cref{thm:anderson} is indeed a special case of \Cref{thm:main_result}. Another open problem is to somehow strengthen \Cref{thm:main_result} in a natural way to make it applicable for CRNs such as the one presented in \Cref{ex:no_converse}.

A further research direction is to incorporate probabilities. One may associate a probability to each T-invariant by multiplying the probabilities of the corresponding reactions. An open problem is to find a probabilistic version of \Cref{thm:main_result} to make predictions about long-term behavior of probabilistic computational models of CRNs, such as the models of \cite{ProgrCRNs/winfree_solo,DBLP:conf/dna/CummingsDS14,DBLP:journals/nc/SoloveichikCWB08}.

\subsection*{Acknowledgements}
We thank David Anderson for kindly explaining his work during the Banff International Research Station (BIRS) workshop on CRNs (14w5167). Also, we thank the organizers of this workshop during which this research was initiated. We are indebted to Matthew Johnston for carefully reading an earlier version of this paper and for providing useful comments. And in particular for finding a counterexample to a conjecture in an earlier version of this paper. We finally thank the five referees for their useful comments. R.B.\ is a postdoctoral fellow of the Research Foundation -- Flanders (FWO).

\bibliographystyle{abbrv}
\bibliography{crns_petri}

\begin{thebibliography}{10}

\bibitem{Anderson/CRN/Domination}
D.~F. Anderson, G.~A. Enciso, and M.~D. Johnston.
\newblock Stochastic analysis of biochemical reaction networks with absolute
  concentration robustness.
\newblock {\em Journal of The Royal Society Interface}, 11(93), 2014.
\newblock Supporting online material available.

\bibitem{ArisCRN}
R.~Aris.
\newblock Prolegomena to the rational analysis of systems of chemical
  reactions.
\newblock {\em Archive for Rational Mechanics and Analysis}, 19(2):81--99,
  1965.

\bibitem{Boucherie/ClosedTInvariant}
R.~J. Boucherie and M.~Sereno.
\newblock On closed support {T}-invariants and the traffic equations.
\newblock {\em Journal of Applied Probability}, 35(2):473--481, 1998.

\bibitem{DBLP:conf/dna/ChenDS12}
H.-L. Chen, D.~Doty, and D.~Soloveichik.
\newblock Deterministic function computation with chemical reaction networks.
\newblock In D.~Stefanovic and A.~J. Turberfield, editors, {\em Proceedings of
  the 18th International Conference on DNA Computing and Molecular Programming
  (DNA 18)}, volume 7433 of {\em Lecture Notes in Computer Science}, pages
  25--42. Springer, 2012.

\bibitem{ProgrCRNs/winfree_solo}
M.~Cook, D.~Soloveichik, E.~Winfree, and J.~Bruck.
\newblock Programmability of chemical reaction networks.
\newblock In A.~Condon, D.~Harel, J.~N. Kok, A.~Salomaa, and E.~Winfree,
  editors, {\em Algorithmic Bioprocesses}, Natural Computing Series, pages
  543--584. Springer Berlin Heidelberg, 2009.

\bibitem{DBLP:conf/dna/CummingsDS14}
R.~Cummings, D.~Doty, and D.~Soloveichik.
\newblock Probability 1 computation with chemical reaction networks.
\newblock In S.~Murata and S.~Kobayashi, editors, {\em Proceedings of the 20th
  International Conference on DNA Computing and Molecular Programming (DNA
  20)}, volume 8727 of {\em Lecture Notes in Computer Science}, pages 37--52.
  Springer, 2014.

\bibitem{Feinberg/deficiency}
M.~Feinberg.
\newblock Complex balancing in general kinetic systems.
\newblock {\em Archive for Rational Mechanics and Analysis}, 49(3):187--194,
  1972.

\bibitem{FeinbergHorn/deficiency}
M.~Feinberg and F.~Horn.
\newblock Chemical mechanism structure and the coincidence of the
  stoichiometric and kinetic subspaces.
\newblock {\em Archive for Rational Mechanics and Analysis}, 66(1):83--97,
  1977.

\bibitem{Horn/deficiency}
F.~Horn.
\newblock Necessary and sufficient conditions for complex balancing in chemical
  kinetics.
\newblock {\em Archive for Rational Mechanics and Analysis}, 49(3):172--186,
  1972.

\bibitem{DBLP:journals/fuin/MairesseN10}
J.~Mairesse and H.~Nguyen.
\newblock Deficiency zero {P}etri nets and product form.
\newblock {\em Fundamenta Informaticae}, 105(3):237--261, 2010.

\bibitem{DBLP:conf/ac/MemmiR75}
G.~Memmi and G.~Roucairol.
\newblock Linear algebra in net theory.
\newblock In W.~Brauer, editor, {\em Net Theory and Applications}, volume~84 of
  {\em Lecture Notes in Computer Science}, pages 213--223. Springer, 1975.

\bibitem{Oxley/MatroidBook-2nd}
J.~Oxley.
\newblock {\em Matroid theory, Second Edition}.
\newblock Oxford University Press, 2011.

\bibitem{DiscCRN:LCNrecurr}
L.~Paulev\'e, G.~Craciun, and H.~Koeppl.
\newblock Dynamical properties of discrete reaction networks.
\newblock {\em Journal of Mathematical Biology}, 69(1):55--72, 2014.

\bibitem{DBLP:conf/ac/1996petri1}
W.~Reisig and G.~Rozenberg, editors.
\newblock {\em Lectures on Petri Nets {I:} Basic Models}, volume 1491 of {\em
  Lecture Notes in Computer Science}. Springer, 1998.

\bibitem{Shinar12032010}
G.~Shinar and M.~Feinberg.
\newblock Structural sources of robustness in biochemical reaction networks.
\newblock {\em Science}, 327(5971):1389--1391, 2010.

\bibitem{DBLP:journals/nc/SoloveichikCWB08}
D.~Soloveichik, M.~Cook, E.~Winfree, and J.~Bruck.
\newblock Computation with finite stochastic chemical reaction networks.
\newblock {\em Natural Computing}, 7(4):615--633, 2008.

\end{thebibliography}

\end{document}